\documentclass[conference]{IEEEtran}
\IEEEoverridecommandlockouts
\usepackage{graphicx} 
\usepackage{color}
\usepackage{hyperref}
\usepackage[letterpaper, margin=1in]{geometry}

\usepackage{float} 

\begin{document}

\title{
Data-Driven Assessment of Vehicle-to-Grid Capabilities in Supporting Grid During Emergencies: Case Study of Travis County, TX}

\author{\IEEEauthorblockN{Kelsey Nelson}
\IEEEauthorblockA{\textit{Civil, Architectural and Env. Eng.} \\
\textit{The University of Texas at Austin}\\
Austin, USA \\
nelson.n@utexas.edu}

\and
\IEEEauthorblockN{Javad Mohammadi}
\IEEEauthorblockA{\textit{Civil, Architectural and Env. Eng.} \\
\textit{The University of Texas at Austin}\\
Austin, USA \\
javadm@utexas.edu }

\thanks{
This research is supported by the National Defense Science and Engineering Graduate (NDSEG) Fellowship program and the Air Force Office of Scientific Research, Award Numbers: FA9550-24-1-0099 and FA9550-23-1-0203
}
}

\maketitle

\begin{abstract} 
As extreme weather events become more common and threaten power grids, the continuing adoption of electric vehicles (EVs) introduces a growing opportunity for their use as a distributed energy storage resource. This energy storage can be used as backup generation through the use of vehicle-to-grid (V2G) technology, where electricity is sent back from EV batteries to the grid \cite{inproceedings}. With enough participation from EV owners, V2G can mitigate outages during grid emergencies. In order to investigate a practical application of V2G, this study leverages a vast array of real-world data, such as survey results on V2G participation willingness, historical outage data within ERCOT, current EV registrations, and demographic data. This data informs realistic emergency grid scenarios with V2G support using a synthetic transmission grid for Travis County. The results find that as EV ownership rises in the coming years, the simultaneous facilitation of bidirectional charging availability would allow for V2G to play a substantial role in preventing involuntary load shed as a result of emergencies like winter storms. 

\end{abstract}

\section{Introduction}
\subsection{Background and Motivation}

Due to ongoing climate change, the increased frequency and severity of extreme weather events has brought into question the ability of utilities to reliably provide for their customers under extreme conditions \cite{Specturm1} \cite{TexasWeather}. This was exemplified in Texas in February of 2021 during Winter Storm Urie, where 69\% of the state's residents lost power at some point during the storm, with many outages lasting several hours or days \cite{UrieSummary}. 

Meanwhile, Texas has the third highest number of registered electric vehicles (EVs) in the United States, providing an opportunity for residents to personally assist utilities during peaking events by sending electricity back to the grid using vehicle-to-grid technology (V2G) \cite{Registrations}. While vehicle-to-grid (V2G) is still in its infancy in the United States \cite{Scorecard}, other countries have successfully implemented V2G programs for their residents \cite{V2GAvailability}. Understanding what to expect from EV presence and V2G participation rates will be pertinent for city planners looking to understand the role that V2G can play in improving grid resiliency and what policy measures could best support this role, such as subsidizing the installation of bidirectional chargers.

\begin{figure}
\centering
\includegraphics[scale=.3]{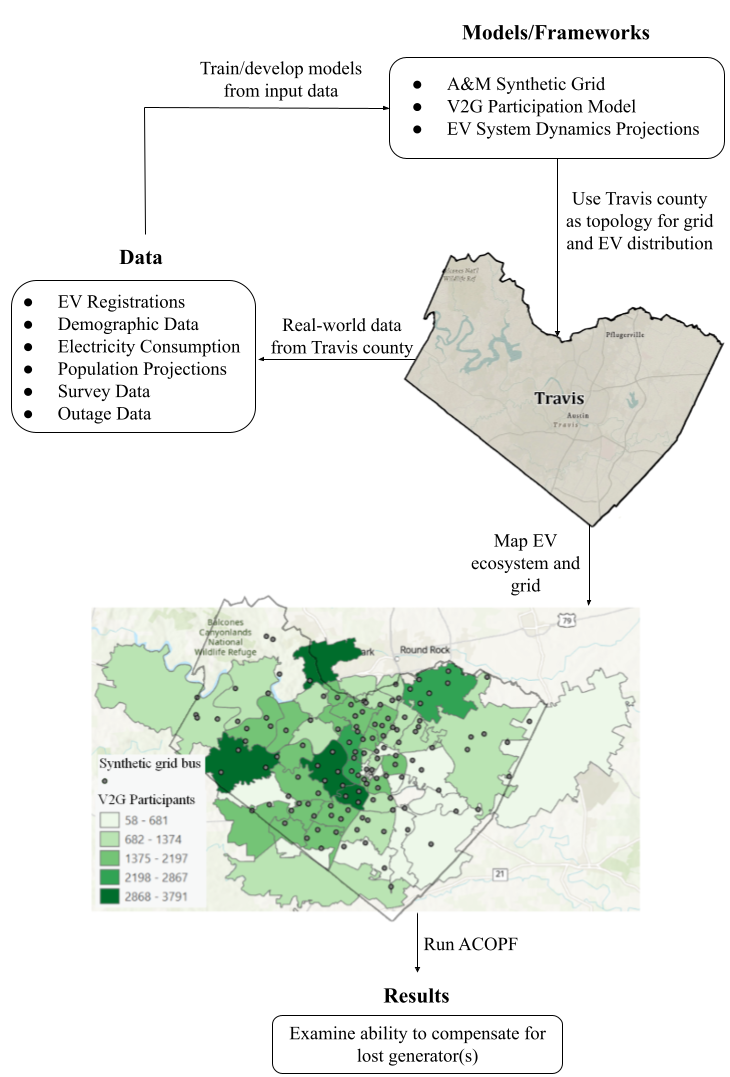}
\caption{An overview of how the proposed framework infuses geography and real-world data from Travis County (in Texas) to create a realistic assessment of how V2G could assist a grid under emergency conditions. The proposed framework consumes a wide range of multi-modal information from EV owners and grid conditions.}
\label{DescriptiveFigure}
\vspace{-.3cm}
\end{figure}

This work uses Travis County (in Texas) as a case study for its application of a V2G simulation framework informed by real-world data. It draws from input data for EV registrations, regional demographics, electricity consumption, population projections, survey results, and historical outage data. These sources inform the models and frameworks applied to the Travis County area, which are a synthetic test grid, a model for projected V2G-participant distribution, and a system dynamics model to project EV presence. The V2G participants are incorporated as generators across the Travis county test grid, where an ACOPF (AC optimal power flow) is run for different outage scenarios. This provides key insights into how facilitating V2G participation can benefit the grid during emergency scenarios. An overview of the architecture of how this case study's data, models, and geography work together to provide its results are shown in Figure \ref{DescriptiveFigure}.

\subsection{Literature Review}

Recently there has been much interest shown in the promise of V2G for grid support. For example, \cite{en16052172} evaluates the ability of managing EV charging and discharging to provide peak shaving and valley filling during day-to-day operations within a distribution testbed. The authors of \cite{pr11082256} similarly investigate peak shaving and valley filling, but within a case study in northern China. Other case studies, such as \cite{ELKHOLY2024118032}, conduct a techno-economic analysis on a microgrid incorporating EV battery storage, demonstrating ``unparalleled performance in terms of both cost-effectiveness and reliability".

In addition to the long term integration of V2G to assist day-to-day grid operations, V2G has also been investigated as a way to assist during emergencies. In \cite{9949967}, V2G is investigated as part of a post-earthquake recovery strategy at the distribution level on a IEEE test grid, verifying its ability to offer effective auxiliary support. Furthermore, \cite{YANG20242786} finds that V2G can support grids through voltage regulation, power angle stability, and harmonic filtering.

\subsection{Contributions and Paper Structure}

The proposed framework, shown in Figure \ref{DescriptiveFigure}, infuses a wide range of multi-modal, regionally specific, information from EV owners, society, and the grid to establish a realistic connection between V2G and power grid resilience. Our study supplements the existing literature by examining how aggregated EVs can contribute at the transmission level during a wide-spread grid emergency. 


The rest of the paper is structured as follows: Section \ref{Data} details the data that this study draws from. Section \ref{Methodology and Key Assumptions} describes how the data is utilized and highlights the key assumptions made for the study's framework and analyses. Next, the results are presented and discussed in Section \ref{Results} and lastly, the study's key takeaways and future works are discussed in Section \ref{Conclusions}.

\section{Multi Modal Data Sets}
\label{Data}

The test bed used for the baseline grid topology and ordinary operating conditions is a 173 bus system developed as part of Texas A\&M University's Electric Grid Test Case Repository \cite{TestGrid}. This grid was created to serve as a realistic model of a transmission grid across the geographical bounds of Travis County. In order to appropriately simulate a winter grid emergency, data was used from reports on the impacts of Winter Storm Urie on generation capacity and demand within ERCOT \cite{2021Report}, \cite{ERCOTReport}. 


For the EVs across Travis County, EV Atlas's State EV registration database \cite{Registrations} was used. These current registrations were then projected into future years using growth scenarios described in Section \ref{Methodology and Key Assumptions}.
For the percentage of residents willing to participate in V2G for grid emergencies, results from a survey on public willingness to participate in bidirectional charging programs were used \cite{doi:10.1177/03611981241253608}. These responses trained a predictive model which was applied to a synthetic population created with data from the American Communities Survey \cite{ACS} and public use microdata sample areas as inputs to a program, PopGen 2.0 from the Mobility Analytics Research Group \cite{PopGen}.


\section{Methodology and Key Assumptions}
\label{Methodology and Key Assumptions}

The raw data from \cite{doi:10.1177/03611981241253608} was used to train a model to identify demographic predictors for an individual's likelihood of supplying electricity via V2G during grid emergencies. Each response was encoded on a numerical scale of 1-5 for a linear regression model.


Next, a synthetic population was created using PopGen 2.0 with input data from the 2021 American Communities Survey and public use micro data samples. The linear regression model was used on the synthetic population in order to predict the overall percentage of residents in each zip code that would respond ``I definitely would not participate, ``I probably would not participate", ``I am unsure", ``I probably would participate", or ``I definitely would participate" by rounding to the nearest corresponding value on the likert scale. Each response category was assigned an approximated participation rate as shown in table Tab.~\ref{Likert}.

\begin{table}[htbp]
\caption{Model numerical result and corresponding survey response and participation rates} 
\label{Likert}
\begin{center}
\begin{tabular}{|p{1cm}|p{4cm}|p{1.5cm}|}

\hline
Model Value & Survey Response & Participation Rate\\
 \hline
 1  & I definitely would not participate & 0\% \\
 \hline
 2 & I probably would not participate & 25\% \\
 \hline
 3 & I am unsure & 50\% \\
  \hline
 4 & I probably would participate\% & 75\% \\
 \hline
 5 & I definitely would participate\% & 100\% \\
 \hline
 \end{tabular}
 \end{center}
\end{table}

With these responses, an overall percentage of residents expected to contribute with V2G during a grid emergency was found for each zip code in the test grid. This percentage was applied to actual EV registrations to create a map of V2G participants across the test grid \cite{Registrations}.

The V2G participants are scaled to EV fleet projections for the years 2030, 2035, and 2040 by using EV market shared benchmarks \cite{10123430}. This was done using a system dynamics model. Within the model, each zip code's population is assumed to scale proportionally with population projections for Austin \cite{Robinson}. The average lifetime of a light duty vehicle is assumed to be 15 years \cite{Keith_2019} the percentage of residents who own cars doesn't change throughout the model’s runtime \cite{TXDOT2023}. The model uses estimates for business as usual year-on-year market share changes from \cite{NBERw28933} and accounts for how EV adoption incentives will influence these year-on-year changes by creating a market share multiplier \cite{Narassimhan_2018}.



Projections for EV registrations and V2G participation rates were mapped across zip codes to be added as additional generation capacity within the Texas A\&M's Travis county test grid \cite{TestGrid}. Each vehicle's power output is assumed to be 7kW/vehicle to simulate the typical speed of a level 2 charger in North America \cite{Level2Speed}. Demand across the test grid was scaled to simulate a winter peak using Austin Energy's reported winter 2021 peak demand \cite{AE2021}. 



\begin{figure} [htbp]
\centering
\includegraphics[scale=.5]{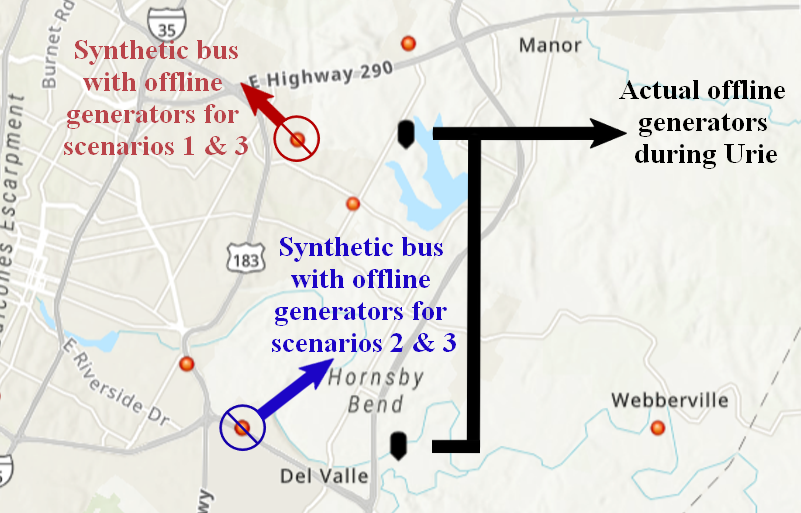}
\caption{This map displays the actual location of buses where generators were taken offline during Winter Storm Urie (in black) and the synthetic grid's generators that were taken offline for this study (in red and blue).}
\label{Generators}
\end{figure}

The actual generators that went offline within Travis county during Winter Storm Urie were mapped against the A\&M test grid in order to take generators with a similar capacity, fuel type, and geographical location offline. This is shown in Figure \ref{Generators}. The following three grid emergency scenarios are considered: (1) The natural gas power plants at bus 10 were taken offline, removing 930 MW of capacity, (2) The natural gas power plants at bus 172 were taken offline, removing 639 MW of capacity, and (3) The natural gas power plants at both bus 10 and 172 were taken offline, removing 1569 MW of capacity.

An ACOPF was run by importing grid elements into a Python package called Pandapower \cite{pandapower.2018} for each outage scenario with the following V2G participation levels: No participation, participation with 2025 EV registrations, participation with 2030 EV registrations, participation with 2035 EV registrations, and participation with 2040 EV registrations. Another key assumption of this study is that all willing V2G participants have access to a bidirectional charger in order to support the grid. While this is not currently the reality for EV owners in Travis County, this assumption is necessary to evaluate the benefits of accessibility to bidirectional chargers for emergency situations such as those simulated in this study. 

Lastly, because the EV registrations are available by make and model, each registered vehicle's range was cataloged by referencing manufacturer websites. These ranges were used later in the analysis to visualize the maximum percentage of the V2G fleet with remaining energy to supply the grid with as the grid emergency continues. 

\section{Results and Discussion}
\label{Results}

The predictive model for V2G participation used the following feature variables: age, sex, income, and education level. On the testing split of the survey data, it demonstrated robust performance with an $R^2$  value of .79. $R^2$ ranges from 0 to 1 and describes the strength of the linear relationship between a function and its dependent variables, with a value of 1 denoting the strongest possible relationship \cite{R2}. The mean absolute error was relatively low at .56, meaning that on average the predicted answer was on or adjacent to the respondent's actual answer on the 5-point likert scale.

\begin{figure} [htbp]
\centering
\includegraphics[scale=.6]{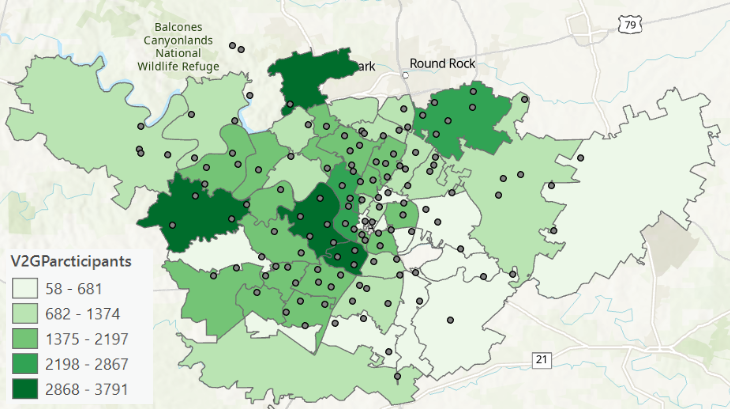}
\caption{Spatial distribution of V2G participants within Travis County test grid.}
\label{Map}
\end{figure}

After applying this model to a synthetic population across Travis county, the percentage of people within each zip code expected to participate in V2G for a grid emergency was applied to the number of registered EVs within each zip code. This distribution of V2G participants is mapped against the synthetic grid's substations in Figure \ref{Map}.

With the added generation capacity from V2G participants, an ACOPF was run on the three winter emergency scenarios obtained in Section \ref{Methodology and Key Assumptions} (scenario 1 takes the generators at bus 10 offline, scenario 2 takes the generators at bus 172 offline, and scenario 3 takes the generators at both bus 10 and bus 172 offline). For cases where the ACOPF did not converge due to insufficient generation capacity, the feasibility gap is evaluated by showing the unmet electricity demand as a percentage of total load. This percentage can be considered to be proximal to the amount of involuntary load shed, or percentage of residents without power, under each scenario. 

\begin{table}[htbp]
\caption{ACOPF results by outage scenario} 
\label{ACOPF Results}
\begin{center}
\begin{tabular}{|c|c|c|c|c|c|}
\hline
&
\multicolumn{5}{|p{5cm}|}{Involuntary load shed as a percentage of system-wide demand} \\
\hline
Scenario & No V2G & 2025 & 2030 & 2035 & 2040 \\
\hline
1 & 40.7\% & 38.1\% & 22.5\% & 3.47\% & 0.00\%\\
\hline
2 & 34.7\% & 32.1\% & 13.2\% & 0.00\% & 0.00\%\\
\hline
3 & 53.8\% & 51.3\% & 35.4\% & 16.4\% & 0.00\%\\
\hline
 \end{tabular}
 \end{center}
\end{table}

All scenarios converge when projecting the registered EV fleet to 2040 levels. Scenario  2 converges before other scenarios at 2035 levels. Even in scenarios that don't converge, there is still a substantial amount of involuntary load shed avoided by the aggregated participation of V2G across the county, seen by comparing each column's load shed to that of the ``No V2G" column. 

\begin{figure} [htbp]
\centering
\includegraphics[scale=.6]{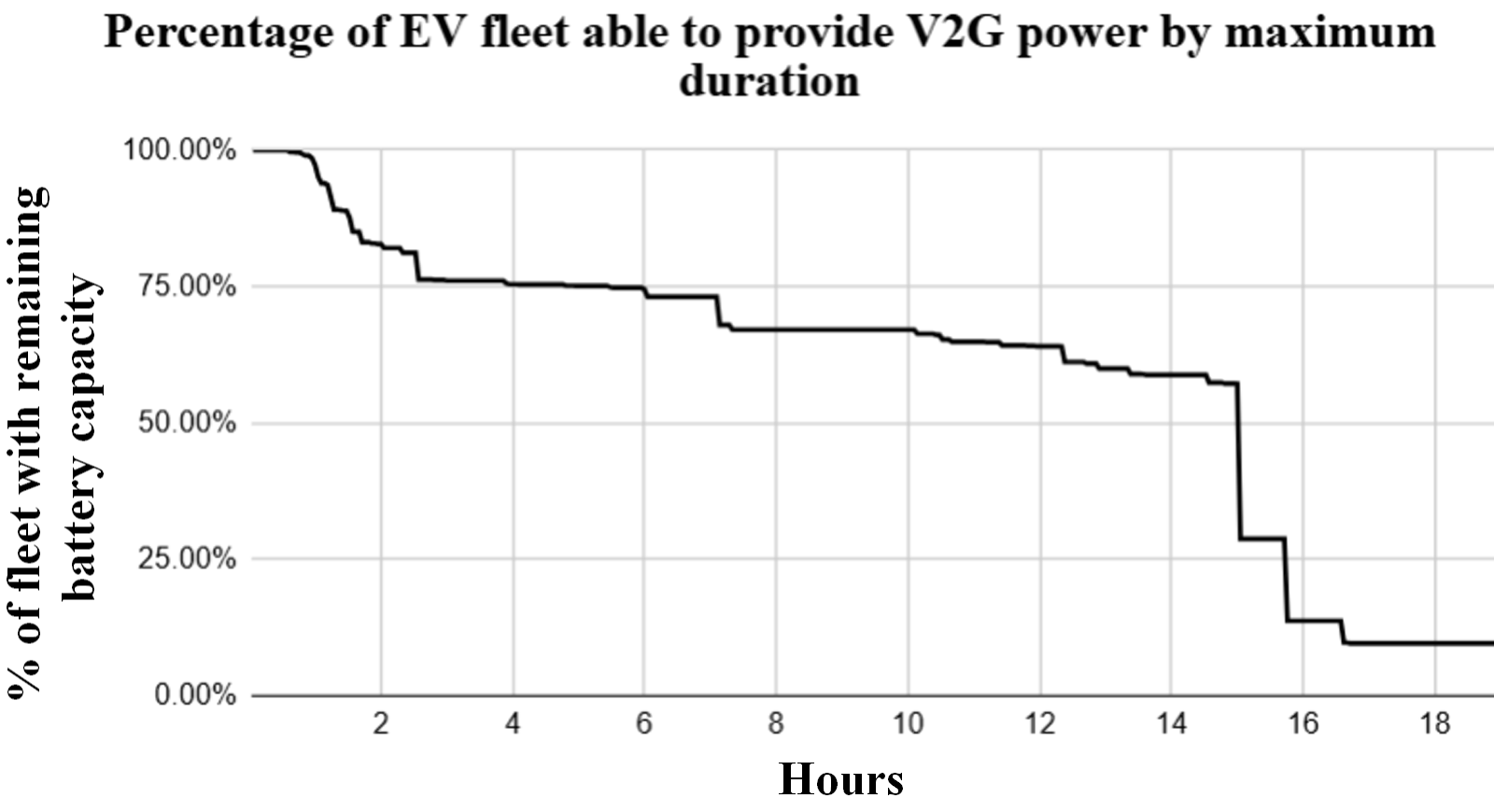}
\caption{This figure shows how over time, V2G participants batteries will become depleted if all participating EVs dispatch power through a level 2 charger at time = 0.}
\label{Battery}
\end{figure}

The time that the EV fleet will be able to sustain the grid at this level, however, is limited by battery capacity. Figure \ref{Battery} shows what percentage of EVs would be able to continue providing V2G support as the emergency event continues over time with the current makeup of battery ranges. 

Figure \ref{Battery} shows how EVs across the county would lose the ability to contribute over time due to having a depleted battery through sending electricity back to the grid. Figure \ref{Battery} uses battery ranges for the current makeup of the EV fleet, many of which  (about 25\%) are plug-in hybrid electric vehicles. Currently, some EV adopters prefer plug-in hybrids over fully electric vehicles due to range anxiety and lack of charging availability \cite{RangeAnxiety}. As EV adoption continues, their ranges evolve, and the charging infrastructure landscape shifts, this fleet makeup may also shift, impacting the duration that V2G participants can be expected contribute to the grid \cite{HybridMarket}. With the current fleet, however, a substantial portion (over half) would be able to contribute to the grid for over 12 hours, showing promise for the fleet's ability to meaningfully assist in an extended grid emergency even when accounting for hybrid vehicles with relatively low battery capacity. 

\section{Conclusions}
\label{Conclusions}

By using real-world EV data, informed estimates for V2G participation, mimicking real-world outage scenarios, and using a geographically appropriate grid topology, this study finds that V2G has the ability to substantially assist the grid during an emergency. This suggests that the deployment of V2G infrastructure will be a highly valuable tool for utilities looking to take advantage of EVs as a growing backup generation resource. 

Future works can address some of the limitations of this preliminary study and expand on its results. One way to do this would be to examine different discharge rates, such as using DC fast charging to increase the EV fleet's power output. Furthermore, given the small battery capacity of some plug-in hybrids, V2G from these EVs may not be as beneficial for outages lasting several hours or even days. Given their suitability for short term emergencies, however, it will also be of great interest to use the framework of this case study as the basis for a techno-economic analysis on how investing in V2G infrastructure can benefit utilities over time by providing access to peak shaving and emergency backup power services.

\bibliography{main.bib}
\bibliographystyle{IEEEtran}

\end{document}